\documentclass[10pt]{article}
\usepackage{setspace}
\usepackage{afterpage}
\usepackage{amsmath}
\usepackage{ltablex}
\usepackage{longtable}
\usepackage{amssymb}
\usepackage{graphicx}
\usepackage{multirow}
\usepackage{multicol}
\usepackage{textcomp}
\usepackage[utf8]{inputenc}
\usepackage[british]{babel}
\usepackage{float}
\usepackage{gensymb}
\usepackage{amsfonts}
\usepackage{cite}
\usepackage{makecell}
\usepackage{caption}
\usepackage{epstopdf}
\usepackage{multicol}
\usepackage[export]{adjustbox}
\usepackage[paperwidth=595.44pt,
paperheight=841.68pt,top=72pt,right=72pt,bottom=72pt,
left=72pt]{geometry
}
\begin{document}
\title{Controlled bidirectional quantum teleportation of superposed coherent state using five qubit cluster-type entangled coherent state as a resource
}
\date{}
\maketitle
\begin{center}
Ravi Kamal Pandey\textsuperscript{$\alpha$},
Phool Singh Yadav\textsuperscript{$\beta$}
Ranjana Prakash\textsuperscript{$\gamma$},
Hari Prakash\textsuperscript{$\delta$},
\\
\bigskip
\ Physics Department, University of Allahabad, Allahabad- 211002, India.
\\
E-mail: $\alpha$ ravikamalpandey@gmail.com\\
$\beta$ phsyadav@rediffmail.com\\
$\gamma$ prakash\_ranjana1974@rediffmail.com\\
$\delta$ prakash\_hari123@rediffmail.com
\end{center}

\begin{abstract}
We propose a scheme of bi-directional quantum teleportation of coherent state qubit between two distant partners Alice and Bob, with the consent of controller, Charlie. We use five-mode cluster-type entangled coherent state as the quantum resource to achieve this task. The scheme uses linear optical devices such as beam splitter, phase shifters, and photon counters. It is shown that for moderately large coherent amplitude, near perfect bi-directional controlled teleportation can be obtained in terms of the average fidelity of teleportation.
\end{abstract}
\section{Introduction}
Quantum teleportation allows a sender (Alice) to communicate to a remote receiver (Bob) by encoding information in quantum state of a system \cite{bennett1993teleporting}. Most often, it is required that both Alice and Bob can communicate simultaneously which (in most simplistic scenario) can be acheived using two quantum teleportation protocol, one from Alice to Bob and vice-versa. This shall require two Bell states being shared between Alice and Bob and both have to make a Bell state measurement (BSM) on their location, subsequently, sharing classically the BSM outcome so as to make appropriate unitary operation and retrieve the original state. However, in such a protocol, if one of the partner denies to make BSM, even then single way communication is still possible. One can circumvent such issue by considering Zha et al.f \cite{zha2013bidirectional} scheme of controlled bi-directional quantum teleportation (CBQT). The scheme uses a genuinely entangled five-qubit cluster entangled state, instead of tensor product of two Bell state, as quantum resource to obtain two-way communication between Alice and Bob with the consent of controller, Charlie. In fact, it has been shown by Shukla et al. \cite{shukla2013bidirectional} that there is nothing special about five-qubit cluster state, and that one can construct infinite number of five-qubit genuinely entangled state which can be used to obtain CBQT. Following these, numerous schemes of bidirectional teleportaion of single-qubit as well as multi-qubits were proposed using genuinely entangled resource of five or more qubits\cite{li2013bidirectional,yan2013bidirectional,
chen2014bidirectional,zhang2015bidirectional,
zhang2015bidirectiona,sang2016bidirectional,
li2016bidirectional,choudhury2016bidirectional,
zhou2019bidirectional,verma2020comment,
verma2020bidirectiona,verma2020bidirection,
verma2020bidirectional1,huo2021controlled,
verma2021bidirectional}. 

The coherent state of radiation field can be a promising physical system for performing many important quantum information tasks\cite{sanders2012review}. Phase opposite coherent state of equal amplitude has been used to encode single qubit information which can be teleported using entangled coherent state (ECS) as quantum resource\cite{van2001entangled}. One can even obtain teleportation fidelity almost unity using only linear optics and photon number resolving detectors\cite{prakash2007improving}. Variants of quantum teleportation as well as other quantum information processing can also be obtained using ECS and linear optics with almost perfect success
\cite{prakash2009increase,prakash2009swapping,prakash2009quantum,
prakash2009entanglement,prakash2010improving,prakash2010almost,
mishra2010teleportation,prakash2019controlled,
pandey2019controlled,pandey2021high}.

Recently, an attempt has been made to obtain bi-directional teleportation of information encoded in superposition of phase opposite coherent state (SCS) using 3-mode entangled coherent state as a resource\cite{aliloute2021bidirectional}. The success probability of the scheme was shown to be $1/2$. However, there are few lapses in their proposed scheme. For example, the mixing of modes using a beam splitter has been done for modes which are at different location, i.e, one mode is with Alice and the other mode is with Bob.  Such non-local mixing is not allowed unless we have some other ancillary mode and using teleportation. Furthermore, the two controlled SWAP gate-operations performed by Alice and Bob are conditioned to the state of the remaining mode with them. However, no corresponding change is considered by the authors in the state of the system after making the measurement. Therefore, such a scheme of obtaining bi-directional teleportation of SCS is not feasible.  

We present a scheme of obtaining CQBT using five-mode cluster-type genuine ECS in this paper. Our scheme requires linear optical devices, such as symmetric beam splitter, phase shifter and photon resolving detectors for its implementation. The average fidelity of teleportation becomes almost unity for appreciable mean coherent amplitude of the involved modes.

\section{BCQT scheme}

\begin{figure}[t]
  \includegraphics[width=\linewidth]{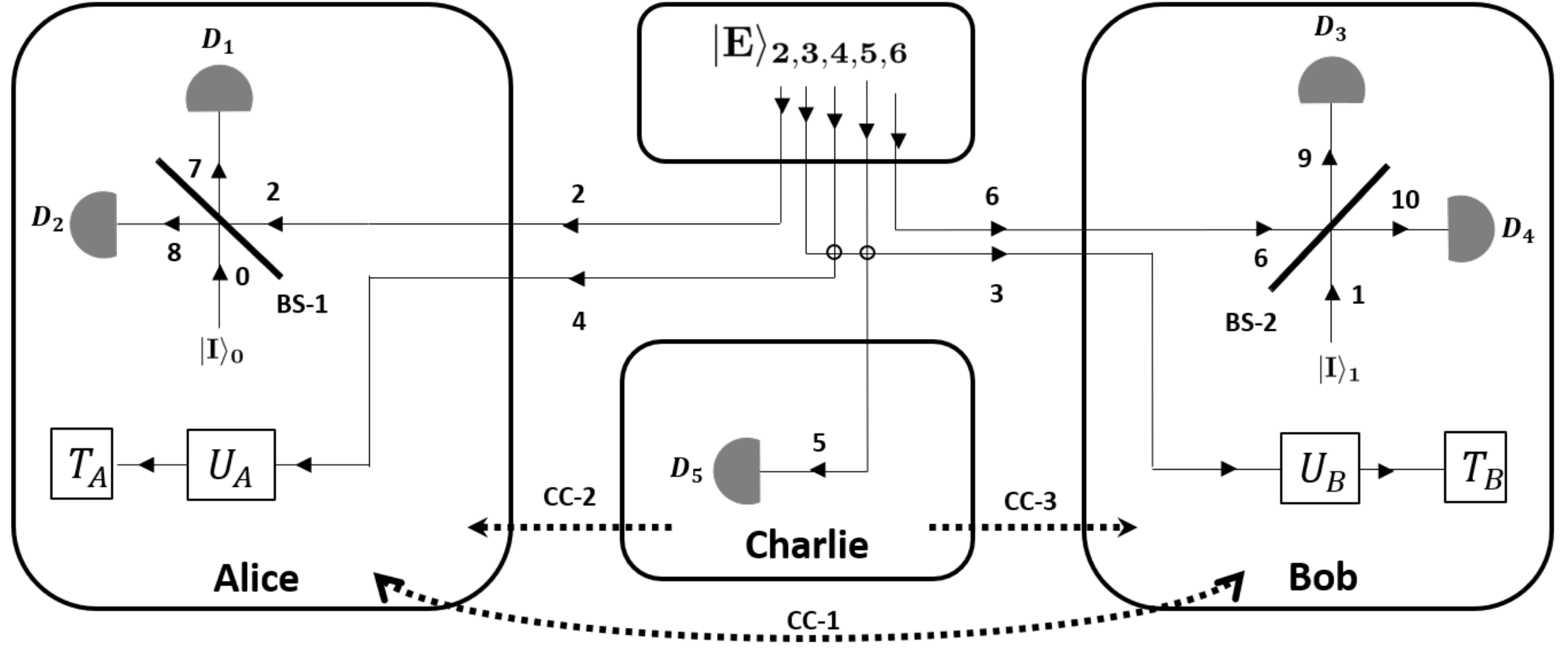}
  \caption{Schematic of the proposed scheme for the BCQT of SCS using five mode cluster-type ECS. Alice mixes the information state, $|I\rangle_{0}$ in mode 0 with mode 1 of entangled state $|E\rangle_{2,3,4,5,6}$ using symmetric beam splitter (BS-I) and performs photon counting measurement on the output modes 7 and 8 using detectors $D_1$ and $D_2$, respectively. Bob also mixes his information state, $|I\rangle_{1}$, with mode 6 using symmetric beam splitter (BS-II) and perform PC on output modes 9 and 10 using detectors $D_3$ and $D_4$. Both Alice and Bob communicates their PC outcome to each other using classical channel, CC-1. The controller, Charlie, does a PC on mode 5 using detector $D_{5}$, and communicaate the outcome to Alice and Bob using classical channels CC-2 and CC-3, respectively. Depending on the classical inputs, Alice and Bob does appropriate unitary operation, $U_{A}$ and $U_{B}$, on modes 4 and 3 respectively, which completes the teleportation protocol.}
  \label{fig:1}
\end{figure}

We consider three remote partners, Alice, Bob and Charlie. Alice and Bob each has a single qubit information encoded in phase opposite coherent states, $|\alpha\rangle$, $|-\alpha\rangle$ in mode $0$ and $1$ given by
\begin{equation}
\label{eqn:1}
\begin{gathered} 
|I^A\rangle_0=\epsilon_+|\alpha\rangle_0+\epsilon_-|-\alpha\rangle_0
\\
|I^B\rangle_1=\omega_+|\alpha\rangle_1+\omega_-|-\alpha\rangle_1,
\end{gathered} 
\end{equation}
respectively. The normalization of the information states requires,
 $|\epsilon|_{+}^{2}+|\epsilon|_{-}^{2}+2x^{2}Re(\epsilon_{+}^{\star}\epsilon_{-})=|\omega|_{+}^{2}+|\omega|_{-}^{2}+2x^{2}Re(\omega_{+}^{\star}\omega_{-})=1$. The information states can also be written in orthogonal basis using even and odd coherent states \cite{dodonov1974even},
\begin{equation}
\label{eqn:2}  
\begin{gathered}   
        |+,\alpha\rangle=[\sqrt{2(1+x^{2})}]^{-1}(|{\alpha}\rangle+|-{\alpha}\rangle)
       \\
 |-,\alpha\rangle=[\sqrt{2(1-x^{2})}]^{-1}(|{\alpha}\rangle-|-{\alpha}\rangle,
    \end{gathered}
  \end{equation}
as, 
\begin{equation}
\label{eqn:3}  
\begin{gathered}  
        |I^A\rangle_0=A_+|+,\alpha\rangle_0+A_-|-,\alpha\rangle_0 \\
  |I^B\rangle_1=B_+|+,\alpha\rangle_1+B_-|-,\alpha\rangle_1.      
    \end{gathered}
  \end{equation}
Without loss in generality, we can write $A_{+}=\cos \theta /2$, $A_{-}=\exp(i\phi)\sin\theta/2$ and $B_{+}=\cos \theta ^{'} /2$,  $B_{-}=\exp(i\phi^{'})\sin\theta^{'}/2$ . In order to fulfil the task of simultaneously teleporting the information to each other with the consent of Charlie; Alice, Bob and Charlie share a five mode cluster-type ECS given as,
\begin{eqnarray}
\label{eqn:4} 
|E\rangle_{2,3,4,5,6} &=& [2\sqrt{1+2x^6+x^8}]^{-1}(|\alpha,\alpha,\alpha,\alpha,\alpha\rangle+|\alpha,\alpha,-\alpha,-\alpha,-\alpha\rangle+|-\alpha,-\alpha,\alpha,-\alpha,\alpha\rangle
\nonumber \\ && 
+|-\alpha,-\alpha,\alpha,-\alpha,\alpha\rangle)_{2,3,4,5,6}
\end{eqnarray}
where, Alice has modes $2$ and $4$, Bob has modes $3$ and $6$, while Charlie has mode $5$. At this point, the overall state of the system consisting of seven modes can be written as,
\begin{equation}
\label{eqn:5} 
|\psi\rangle_{0,1,2,3,4,5,6}=|I^A\rangle_0|I^B\rangle_1|E\rangle_{2,3,4,5,6}.
\end{equation} 
Alice mixes modes $0$ and $2$ using a symmetric beam splitter (BS-I) which outputs modes $7$ and $8$. Similarly, Bob also mixes modes $1$ and $6$ using another symmetric beam splitter (BS-II) to become modes $9$ and $10$ (see Fig. \ref{fig:1}). Both BS-I and BS-II are fitted with two phase shifters at its second input and second output port. The phase shifter changes state $|\alpha\rangle$ to $|-i\alpha\rangle$. This combination of a beam splitter with two phase shifters transforms the input ports $x$ and $y$, to output modes $u$ and $v$ as, $|{\alpha},{\beta}\rangle_{x,y}\rightarrow |\frac{{\alpha}+{\beta}}{\sqrt{2}},\frac{{\alpha}-{\beta}}{\sqrt{2}}\rangle_{u,v}$. Then, the overall state of the system can be written as,
\begin{eqnarray}
\label{eqn:6} 
|\psi\rangle_{7,8,4,5,3,10,9}&=&[2\sqrt{1+2x^6+x^8}]^{-1}
[\epsilon_+\omega_+
(|\sqrt{2}\alpha,0,\alpha,\alpha,\alpha,\sqrt{2}\alpha,0\rangle
+|\sqrt{2}\alpha,0,-\alpha,-\alpha,\alpha,0,-\sqrt{2}\alpha\rangle
\nonumber \\ && \nonumber
+|0,\sqrt{2}\alpha,\alpha,-\alpha,-\alpha,\sqrt{2}\alpha,0\rangle
+|0,\sqrt{2}\alpha,-\alpha,\alpha,-\alpha,0,-\sqrt{2}\alpha\rangle)_{7,8,4,5,3,10,9}
\\ && \nonumber
+\epsilon_+\omega_-
(|\sqrt{2}\alpha,0,\alpha,\alpha,\alpha,\sqrt{2}\alpha,0\rangle
+|\sqrt{2}\alpha,0,-\alpha,-\alpha,\alpha,-\sqrt{2}\alpha,0\rangle
\\ && \nonumber
+|0,\sqrt{2}\alpha,\alpha,-\alpha,-\alpha,0,\sqrt{2}\alpha\rangle
+|0,\sqrt{2}\alpha,-\alpha,\alpha,-\alpha,0,-\sqrt{2}\alpha\rangle)_{7,8,4,5,3,10,9}
\\ && \nonumber
+\epsilon_-\omega_+
(|0,-\sqrt{2}\alpha,\alpha,\alpha,\alpha,\sqrt{2}\alpha,0\rangle
+|0,-\sqrt{2}\alpha,0,-\alpha,-\alpha,\alpha,-\sqrt{2}\alpha,0\rangle
\\ && \nonumber
+|0,\sqrt{2}\alpha,\alpha,-\alpha,-\alpha,\sqrt{2}\alpha,0\rangle
+|0,\sqrt{2}\alpha,-\alpha,\alpha,-\alpha,0,-\sqrt{2}\alpha\rangle)_{7,8,4,5,3,10,9}
\\ && \nonumber
+\epsilon_-\omega_-
(|\sqrt{2}\alpha,0,\alpha,\alpha,\alpha,\sqrt{2}\alpha,0\rangle
+|\sqrt{2}\alpha,0,-\alpha,-\alpha,\alpha,0,-\sqrt{2}\alpha\rangle
\\ && 
+|0,\sqrt{2}\alpha,\alpha,-\alpha,-\alpha,\sqrt{2}\alpha,0\rangle
+|0,\sqrt{2}\alpha,-\alpha,\alpha,-\alpha,0,-\sqrt{2}\alpha\rangle)_{7,8,4,5,3,10,9}].
\end{eqnarray}

Alice performs photon counting (PC) on modes $7$ and $8$ while Bob does PC on modes $9$ and $10$. Following Prakash et al. \cite{prakash2007improving}, we suppose that the photon counter is sensitive enough to distinguish between a vacuum, non-zero even (NZE) and odd photons. We thus write,
\begin{equation}
\label{eqn:7} 
|\pm\sqrt{2}{\alpha}\rangle = x|0\rangle+\frac{1-x^2}{\sqrt{2}}|{NZE,\sqrt{2}\alpha}\rangle \pm\sqrt{\frac{1-x^4}{2}}|{-,\sqrt{2}\alpha}\rangle,
\end{equation}
where, $|NZE,\sqrt{2}\alpha\rangle=[\sqrt{2}(1-x^{2})]^{-1}(|\sqrt{2}\alpha\rangle+|-\sqrt{2}\alpha\rangle-2x|0\rangle)$ represents the normalized state of non-zero even number of photons. At this point, the role of Charlie becomes crucial. Charlie also performs PC on mode $5$ with him so as to distinguish between an even or odd photon. Using Eq.\ref{eqn:7}, we expand modes 7, 8, 9 and 10 in orthogonal basis \{$|0\rangle$,$|NZE,\sqrt{2}\alpha\rangle$,$|-,\sqrt{2}\alpha\rangle\}$, while we expand rest of the modes 4, 5 and 3 in the basis \{$|+,\alpha\rangle,|-,\alpha\rangle$\} using Eq.\ref{eqn:2} to rewrite state in Eq.\ref{eqn:6} as,
{\allowdisplaybreaks
\begin{eqnarray}
\label{eqn:8} 
&&|\psi\rangle_{7,8,4,5,3,10,9}
\nonumber = 
\frac{2x^2 A_{+} B_{+}}{2(1+x^2)\sqrt{1+2x^6+x^8}}|0,0,0,0\rangle_{7,8,10,9}[(1+x^2)|+,\alpha\rangle_{5} |+,\alpha\rangle_{4}|+,\alpha\rangle_{6}
\nonumber \\ &&
+(1-x^2)|-,\alpha\rangle_{5} |-,\alpha\rangle_{4} |+,\alpha\rangle_{6}]
+\frac{x(1-x^2)}{2\sqrt{(1+x^{2})(1+2x^6+x^8)}}
\{B_{+}|NZE,0,0,0\rangle_{7,8,10,9}
\nonumber \\ &&
[(1+x^2)|+,\alpha\rangle_5 |+,\alpha\rangle_4(A_{+}|+,\alpha\rangle_{6}
+ A_{-}|-,\alpha\rangle_6)+(1-x^2)|-,\alpha\rangle_5 |-,\alpha\rangle_4 (\gamma A_{+}|-,\alpha\rangle_{6}+\delta A_{-}|+,\alpha\rangle_6)]
\nonumber \\ &&
+B_{+}|0,NZE,0,0\rangle_{7,8,10,9}[(1+x^2)|+,\alpha\rangle_5 |+,\alpha\rangle_4(A_{+}|+,\alpha\rangle_{6}
- A_{-}|-,\alpha\rangle_6)+(1-x^2)|-,\alpha\rangle_5 |-,\alpha\rangle_4 
\nonumber \\ &&
(-\gamma A_{+}|-,\alpha\rangle_{6}+\delta A_{-}|+,\alpha\rangle_6)]  
+A_{+}|0,0,NZE,0\rangle_{7,8,10,9}[(1+x^2)|+,\alpha\rangle_5 (B_+|+,\alpha\rangle_4 + B_-|-,\alpha\rangle_4)
\nonumber \\ &&
|+,\alpha\rangle_{6}+(1-x^2)|-,\alpha\rangle_5 (\gamma B_{+}|-,\alpha\rangle_{4}+\delta B_{-}|+,\alpha\rangle_{4})|-,\alpha\rangle_6]  
+A_{+}|0,0,0,NZE\rangle_{7,8,10,9}[(1+x^2)|+,\alpha\rangle_5 
\nonumber \\ &&
(B_+|+,\alpha\rangle_4 - B_-|-,\alpha\rangle_4)
|+,\alpha\rangle_{6}+(1-x^2)|-,\alpha\rangle_5 (\gamma B_{+}|-,\alpha\rangle_{4}-\delta B_{-}|+,\alpha\rangle_{4})|-,\alpha\rangle_6 ]\}
\nonumber \\ &&
+\frac{x\sqrt{(1-x^2)}}{2\sqrt{(1+2x^6+x^8)}}
\{B_{+}|ODD,0,0,0\rangle_{7,8,10,9}[(1+x^2)|+,\alpha\rangle_5 |+,\alpha\rangle_4(\gamma A_{+}|-,\alpha\rangle_{6}
+\delta A_{-}|+,\alpha\rangle_6)
\nonumber \\ &&
+(1-x^2)|-,\alpha\rangle_5 |-,\alpha\rangle_4 (A_{+}|+,\alpha\rangle_{6}+\delta A_{-}|-,\alpha\rangle_6)]
+
B_{+}|0,ODD,0,0\rangle_{7,8,10,9}[(1+x^2)|+,\alpha\rangle_5 |+,\alpha\rangle_4
\nonumber \\ &&
(-\gamma A_{+}|-,\alpha\rangle_{6}
+\delta A_{-}|+,\alpha\rangle_6)+(1-x^2)|-,\alpha\rangle_5 |-,\alpha\rangle_4 (A_{+}|+,\alpha\rangle_{6}-\delta A_{-}|-,\alpha\rangle_6)] 
\nonumber \\ &&
+A_{+}|0,0,ODD,0\rangle_{7,8,10,9}[(1+x^2)|+,\alpha\rangle_5 (\gamma B_{+}|-,\alpha\rangle_{4}+\delta B_{-}|+,\alpha\rangle_4)
|+,\alpha\rangle_6+(1-x^2)|-,\alpha\rangle_5 
\nonumber \\ &&
(B_{+}|+,\alpha\rangle_{4}+\delta + B_{-}|-,\alpha\rangle_4)
|-,\alpha\rangle_6]+
A_{+}|0,0,0,ODD\rangle_{7,8,10,9}[(1+x^2)|+,\alpha\rangle_5 (\gamma B_{+}|-,\alpha\rangle_{4}-\delta B_{-}|+,\alpha\rangle_4)
\nonumber \\ &&
|+,\alpha\rangle_6+(1-x^2)|-,\alpha\rangle_5 (B_{+}|+,\alpha\rangle_{4} - B_{-}|-,\alpha\rangle_4)
|-,\alpha\rangle_6]\}
+\frac{(1-x^2)^{2}}{4\sqrt{2(1+2x^6+x^8)}}
\{ |NZE,0,NZE,0\rangle_{7,8,10,9}
\nonumber \\ &&
[\sqrt{(1+x^2)}|+,\alpha\rangle_5 (B_{+}|+,\alpha\rangle_{4}
 + B_{-}|-,\alpha\rangle_4)( A_{+}|+,\alpha\rangle_{6}
+ A_{-}|-,\alpha\rangle_6)+\sqrt{1-x^2}|-,\alpha\rangle_5 
(\gamma B_{+}|-,\alpha\rangle_{4} 
\nonumber \\ &&
+ \delta B_{-}|+,\alpha\rangle_4)
(\gamma A_{+}|+,\alpha\rangle_{6}
+\delta A_{-}|-,\alpha\rangle_6)]
+
|NZE,0,0,NZE\rangle_{7,8,10,9}[\sqrt{(1+x^2)}|+,\alpha\rangle_5 (B_{+}|+,\alpha\rangle_{4} 
\nonumber \\ &&
- B_{-}|-,\alpha\rangle_4)
( A_{+}|+,\alpha\rangle_{6}
+ A_{-}|-,\alpha\rangle_6)+\sqrt{1-x^2}|-,\alpha\rangle_5 
(\gamma B_{+}|-,\alpha\rangle_{4} 
-\delta B_{-}|+,\alpha\rangle_4)(\gamma A_{+}|+,\alpha\rangle_{6}
\nonumber \\ &&
+\delta A_{-}|-,\alpha\rangle_6)]+
|0,NZE,NZE,0\rangle_{7,8,10,9}[\sqrt{(1+x^2)}|+,\alpha\rangle_5 (B_{+}|+,\alpha\rangle_{4} + B_{-}|-,\alpha\rangle_4)
( A_{+}|+,\alpha\rangle_{6}
\nonumber \\ &&
- A_{-}|-,\alpha\rangle_6)+\sqrt{1-x^2}|-,\alpha\rangle_5 
(\gamma B_{+}|-,\alpha\rangle_{4}
+\delta B_{-}|+,\alpha\rangle_4)(\gamma A_{+}|+,\alpha\rangle_{6}
-\delta A_{-}|-,\alpha\rangle_6)]
\nonumber \\ &&
+|0,NZE,0,NZE\rangle_{7,8,10,9}[\sqrt{(1+x^2)}|+,\alpha\rangle_5 (B_{+}|+,\alpha\rangle_{4} - B_{-}|-,\alpha\rangle_4)
( A_{+}|+,\alpha\rangle_{6}- A_{-}|-,\alpha\rangle_6)
\nonumber \\ &&
+\sqrt{1-x^2}|-,\alpha\rangle_5 
(\gamma B_{+}|-,\alpha\rangle_{4} 
-\delta B_{-}|+,\alpha\rangle_4)(\gamma A_{+}|+,\alpha\rangle_{6}
-\delta A_{-}|-,\alpha\rangle_6)]
+\frac{(1-x^4)}{4\sqrt{2(1+2x^6+x^8)}}
\nonumber \\ &&
\{ |ODD,0,ODD,0\rangle_{7,8,10,9}[\sqrt{(1+x^2)}|+,\alpha\rangle_5 (\gamma B_{+}|-,\alpha\rangle_{4}
 +\delta B_{-}|+,\alpha\rangle_4)(\gamma A_{+}|-,\alpha\rangle_{6}
+ \delta A_{-}|+,\alpha\rangle_6)
\nonumber \\ &&
+\sqrt{1-x^2}|-,\alpha\rangle_5 
(B_{+}|+,\alpha\rangle_{4} +  B_{-}|-,\alpha\rangle_4)( A_{+}|+,\alpha\rangle_{6}
+ A_{-}|-,\alpha\rangle_6)]
+|ODD,0,0,ODD\rangle_{7,8,10,9}
\nonumber \\ &&
[\sqrt{(1+x^2)}|+,\alpha\rangle_5 (\gamma B_{+}|-,\alpha\rangle_{4}
 -\delta B_{-}|+,\alpha\rangle_4)(\gamma A_{+}|-,\alpha\rangle_{6}
+ \delta A_{-}|+,\alpha\rangle_6)+\sqrt{1-x^2}|-,\alpha\rangle_5 
\nonumber \\ &&
(B_{+}|+,\alpha\rangle_{4} -  B_{-}|-,\alpha\rangle_4)( A_{+}|+,\alpha\rangle_{6}+ A_{-}|-,\alpha\rangle_6)]+
|0,ODD,ODD,0\rangle_{7,8,10,9}[\sqrt{(1+x^2)}|+,\alpha\rangle_5 (\gamma B_{+}|-,\alpha\rangle_{4}
\nonumber \\ &&
 +\delta B_{-}|+,\alpha\rangle_4)(\gamma A_{+}|-,\alpha\rangle_{6}
- \delta A_{-}|+,\alpha\rangle_6)+\sqrt{1-x^2}|-,\alpha\rangle_5 
(B_{+}|+,\alpha\rangle_{4} +  B_{-}|-,\alpha\rangle_4)( A_{+}|+,\alpha\rangle_{6} - A_{-}|-,\alpha\rangle_6)]
\nonumber \\ &&
+|0,ODD,0,ODD\rangle_{7,8,10,9}[\sqrt{(1+x^2)}|+,\alpha\rangle_5 (\gamma B_{+}|-,\alpha\rangle_{4}
 - \delta B_{-}|+,\alpha\rangle_4)(\gamma A_{+}|-,\alpha\rangle_{6}
- \delta A_{-}|+,\alpha\rangle_6)
\nonumber \\ &&
+\sqrt{1-x^2}|-,\alpha\rangle_5 
(B_{+}|+,\alpha\rangle_{4} -  B_{-}|-,\alpha\rangle_4)( A_{+}|+,\alpha\rangle_{6} - A_{-}|-,\alpha\rangle_6)]
+\frac{(1-x^2 \sqrt{1-x^{4}})}{4\sqrt{2(1+2x^6+x^8)}}
\nonumber \\ &&
\{ |ODD,0,NZE,0\rangle_{7,8,10,9}
[\sqrt{(1+x^2)}|+,\alpha\rangle_5 (B_{+}|+,\alpha\rangle_{4}
 + B_{-}|-,\alpha\rangle_4)(\gamma A_{+}|-,\alpha\rangle_{6}
+ \delta A_{-}|+,\alpha\rangle_6)
\nonumber \\ &&
+\sqrt{1-x^2}|-,\alpha\rangle_5 
(\gamma B_{+}|-,\alpha\rangle_{4} + \delta B_{-}|+,\alpha\rangle_4)( A_{+}|+,\alpha\rangle_{6}
+ A_{-}|-,\alpha\rangle_6)]
+ |ODD,0,0,NZE\rangle_{7,8,10,9}
\nonumber \\ &&
[\sqrt{(1+x^2)}|+,\alpha\rangle_5 (B_{+}|+,\alpha\rangle_{4}
 - B_{-}|-,\alpha\rangle_4)(\gamma A_{+}|-,\alpha\rangle_{6}
+ \delta A_{-}|+,\alpha\rangle_6)+\sqrt{1-x^2}|-,\alpha\rangle_5 
\nonumber \\ &&
(\gamma B_{+}|-,\alpha\rangle_{4} - \delta B_{-}|+,\alpha\rangle_4)( A_{+}|+,\alpha\rangle_{6}
+ A_{-}|-,\alpha\rangle_6)]
+
|0,ODD,NZE,0\rangle_{7,8,10,9}[\sqrt{(1+x^2)}|+,\alpha\rangle_5 
\nonumber \\ &&
(B_{+}|+,\alpha\rangle_{4}
 + B_{-}|-,\alpha\rangle_4)(-\gamma A_{+}|-,\alpha\rangle_{6}
+ \delta A_{-}|+,\alpha\rangle_6)+\sqrt{1-x^2}|-,\alpha\rangle_5 
( -A_{+}|+,\alpha\rangle_{4} + A_{-}|-,\alpha\rangle_4) 
\nonumber \\ &&
(\gamma B_{+}|-,\alpha\rangle_{6} + \delta B_{-}|+,\alpha\rangle_6)]
+
|0,ODD,0,NZE\rangle_{7,8,10,9}[\sqrt{(1+x^2)}|+,\alpha\rangle_5 (B_{+}|+,\alpha\rangle_{4}
 - \delta B_{-}|-,\alpha\rangle_4)
\nonumber \\ && 
 (-\gamma A_{+}|-,\alpha\rangle_{6}
+ \delta A_{-}|+,\alpha\rangle_6)+\sqrt{1-x^2}|-,\alpha\rangle_5 
(-\gamma B_{+}|-,\alpha\rangle_{4} + \delta B_{-}|+,\alpha\rangle_4)( A_{+}|+,\alpha\rangle_{6}
- A_{-}|-,\alpha\rangle_6)]
\nonumber \\ &&
+|NZE,0,ODD,0\rangle_{7,8,10,9}[\sqrt{(1+x^2)}|+,\alpha\rangle_5 (\gamma B_{+}|-,\alpha\rangle_{4}
 + \delta B_{-}|+,\alpha\rangle_4)( A_{+}|+,\alpha\rangle_{6} + A_{-}|-,\alpha\rangle_6)
\nonumber \\ && 
 +\sqrt{1-x^2}|-,\alpha\rangle_5 
(B_{+}|+,\alpha\rangle_{4} + B_{-}|-,\alpha\rangle_4)( \gamma A_{+}|-,\alpha\rangle_{6}
+ \delta A_{-}|+,\alpha\rangle_6)]+
|NZE,0,0,ODD\rangle_{7,8,10,9}
\nonumber \\ &&
[\sqrt{(1+x^2)}|+,\alpha\rangle_5 (\gamma B_{+}|-,\alpha\rangle_{4}
 - \delta B_{-}|+,\alpha\rangle_4)( A_{+}|+,\alpha\rangle_{6} + A_{-}|-,\alpha\rangle_6)+\sqrt{1-x^2}|-,\alpha\rangle_5 
\nonumber \\ &&
(B_{+}|+,\alpha\rangle_{4} - B_{-}|-,\alpha\rangle_4)( \gamma A_{+}|-,\alpha\rangle_{6}
+ \delta A_{-}|+,\alpha\rangle_6)]
+|0,NZE,ODD,0\rangle_{7,8,10,9}[\sqrt{(1+x^2)}|+,\alpha\rangle_5 
\nonumber \\ &&
(\gamma B_{+}|-,\alpha\rangle_{4}
 + \delta B_{-}|+,\alpha\rangle_4)( A_{+}|+,\alpha\rangle_{6} - A_{-}|-,\alpha\rangle_6)+\sqrt{1-x^2}|-,\alpha\rangle_5 
(B_{+}|+,\alpha\rangle_{4} + B_{-}|-,\alpha\rangle_4)
\nonumber \\ &&
( \gamma A_{+}|-,\alpha\rangle_{6}
- \delta A_{-}|+,\alpha\rangle_6)]
+|0,NZE,0,ODD\rangle_{7,8,10,9}[\sqrt{(1+x^2)}|+,\alpha\rangle_5 (\gamma B_{+}|-,\alpha\rangle_{4}
 + \delta B_{-}|+,\alpha\rangle_4)
\nonumber \\ && 
 ( A_{+}|+,\alpha\rangle_{6} - A_{-}|-,\alpha\rangle_6)+\sqrt{1-x^2}|-,\alpha\rangle_5 
(B_{+}|+,\alpha\rangle_{4} - B_{-}|-,\alpha\rangle_4)( -\gamma A_{+}|-,\alpha\rangle_{6}
+ \delta A_{-}|+,\alpha\rangle_6)]\},
\end{eqnarray}}
where $\gamma = \sqrt{\frac{1-x^2}{1+x^2}}$ and $\delta = \sqrt{\frac{1+x^2}{1-x^2}}$. We shall investigate all possible PC outcomes in the following section.

\section{Investigation of PC cases}
From Eq.\ref{eqn:8}, we observe that one of the mode with Alice (7 or 8) as well as Bob (9 or 10) always gives a zero count while the other mode can give a zero, $NZE$ or odd counts. All of these cases are tabulated (see Table 1 in the Appendix). There are a total of 50 mutually exclusive PC outcomes that can be obtained from our scheme. We find that from a total of 50 cases, 32 cases amounts to bi-directional teleportation with fidelity either unity or close to unity (depending upon mean photon amplitude). Whereas, 16 cases lead to uni-directional teleportation and for remaining two cases, teleportation fails. It shall be useful for the discussions what follow to denote the state teleported to Bob as,
\begin{equation}
\label{eqn:9} 
\begin{gathered}
|B^{0}\rangle\sim A_{+}|+,\alpha\rangle
\\
|B^{1}\rangle\sim A_{+}|-,\alpha\rangle
\\
|B^{2}\rangle=A_{+}|+,\alpha\rangle-A_{-}|-,\alpha\rangle
\\
|B^{3}\rangle \sim \gamma A_{+}|-,\alpha\rangle+ \delta A_{-}|+,\alpha\rangle
\\
|B^{4}\rangle \sim \gamma A_{+}|-,\alpha\rangle- \delta A_{-}|+,\alpha\rangle,
\end{gathered}
\end{equation}
while the state teleported to Alice as,
\begin{equation}
\label{eqn:10} 
\begin{gathered}
|A^{0}\rangle \sim B_{+}|+,\alpha\rangle
\\
|A^{1}\rangle\sim B_{+}|-,\alpha\rangle
\\
|A^{2}\rangle=B_{+}|+,\alpha\rangle-B_{-}|-,\alpha\rangle
\\
|A^{3}\rangle \sim \gamma B_{+}|-,\alpha\rangle+ \delta B_{-}|+,\alpha\rangle
\\
|A^{4}\rangle \sim \gamma B_{+}|-,\alpha\rangle- \delta B_{-}|+,\alpha\rangle,
\end{gathered}
\end{equation} 
(where $\sim$ denotes un-normalized state) and the unitary operations,
\begin{equation}
\label{eqn:11} 
\begin{gathered}
U_{1}=|+,\alpha\rangle\langle -,\alpha|+|-,\alpha\rangle\langle +,\alpha|
\\
U_{2}=|+,\alpha\rangle\langle -,\alpha|-|-,\alpha\rangle\langle +,\alpha|
\\
U_{3}=|+,\alpha\rangle\langle +,\alpha|-|-,\alpha\rangle\langle -,\alpha|.
\end{gathered}
\end{equation} 
For information state $|\psi^I\rangle$ and teleported state $|\psi^T\rangle$, the teleportation fidelity is given by,
\begin{equation}
\label{eqn:12}
F=|\langle \psi^I|\psi^T\rangle|^2.
\end{equation} 
Since, there are two teleportations involved, we have two fidelities as well. For teleportation from Alice to Bob, the fidelity obtained will be one of the following, 
\begin{equation}
F^{AB}_{1}=|A_{+}|^{2}; F^{AB}_{2}=|A_{-}|^{2}; F^{AB}_{3}=1-\frac{x^{4}\sin^{2}\theta}{1+x^{4}-2x^{2}\cos\theta}.
\end{equation}
On the other hand, for teleportation from Bob to Alice, the fidelity can be, 
\begin{equation}
F^{BA}_{1}=|B_{+}|^{2}; F^{BA}_{2}=|B_{-}|^{2}; F^{BA}_{3}=1-\frac{x^{4}\sin^{2}\theta^{'}}{1+x^{4}-2x^{2}\cos\theta^{'}}.
\end{equation}
 We shall now discuss the PC outcomes. 
\subsection{Cases of failure}
\begin{figure}[t]
  \includegraphics[width=\linewidth]{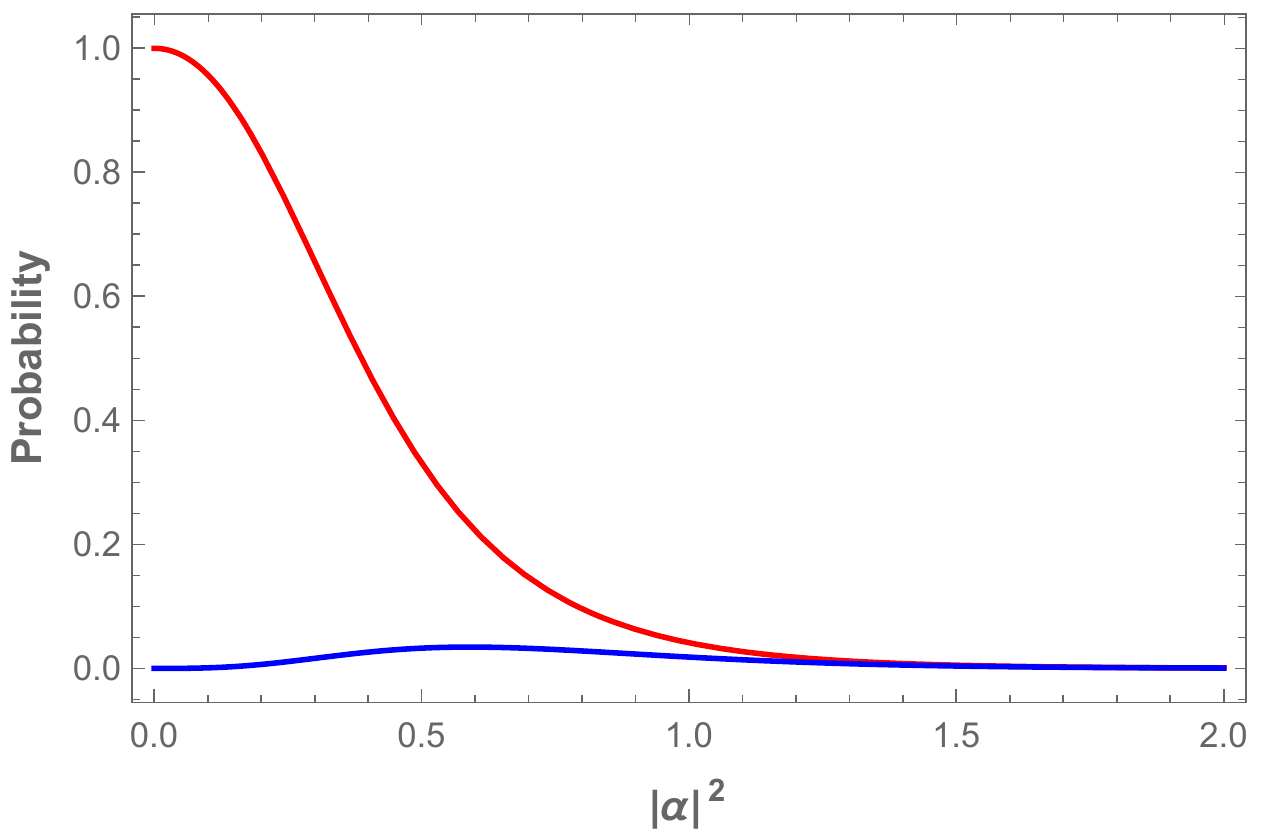}
  \caption{Red and blue curve shows the variation of  $P^{max}_{I,+}$ and $P^{max}_{I,-}$ respectively with respect to mean photons in the coherent state, $|\alpha|^{2}$. Both rapidly decreases with increasing $|\alpha|^{2}$ to become almost zero for $|\alpha|^{2}\geqslant 2$.}
  \label{fig:2}
\end{figure}

These cases correspond to situations when both the output of BS-I and BS-II give vacuum while Charlie obtain either even or odd counts in mode 5. For case 1 and 2, the teleported with 
\{Alice,Bob\} are given by \{$|A^{0}\rangle,|B^{0}\rangle$\} and \{$|A^{1}\rangle,|B^{1}\rangle$\}, respectively, amounting to respective fidelity $\{F_{1}^{BA},F_{1}^{AB}\}$ and $\{F_{2}^{BA},F_{2}^{AB}\}$. For both of these cases, the teleported states can not be transformed to the information states using any unitary transformation. Hence, for these cases, we say that the teleportation fails for both sides. The probability of obtaining such a case are given by
\begin{equation}
\label{eqn:13} 
P_{I,+}=\frac{2x^{4}(1+x^{2})|A_{+}|^{2}||B_{+}|^{2}}{1+2x^{6}+x^{8}}
\end{equation} 
for case 1, and
\begin{equation}
\label{eqn14} 
P_{I,-}=\frac{2x^{4}(1-x^{2})^{3}|A_{+}|^{2}||B_{+}|^{2}}{(1+x^{2})^{2}(1+2x^{6}+x^{8})}
\end{equation}
for case 2. Both $P_{I,+}$ and $P_{I,-}$ depend on information parameters $\theta$ and $\theta^{'}$ as well as $|\alpha|^{2}$. However, both these expressions contain the product of $x^{4}$ with $|A_{+}|^{2}|B_{+}|^{2} $, where the latter is always less than or equal to one, while $x^{4}\simeq 3.3 \times 10^{-4}$ for $|\alpha|^{2}=2$. Hence, for appreciable coherent amplitude ($|\alpha|^{2}\geq 2$) both these probabilities become vanishingly small implying little chance of occurrence of these cases. The probabilities are plotted for the special case when $|A_{+}|^{2}|B_{+}|^{2}=1$, which is the upper bound for probability obtained for any other value of $A_{+}$ and $B_{+}$. Both $P^{max}_{I,+}$ and $P^{max}_{I,-}$ become almost zero for $|\alpha|^{2}\geq 2$ (Fig.\ref{fig:2}).
\subsection{Cases of unidirectional teleportation } 
\subsubsection{3 to 10}
Cases 3 to 10 occur for collective PC outcomes (NZE,0,0,0,even), (NZE,0,0,0,odd), (0,NZE,0,0,even), (0,NZE,0,0,odd), (odd,0,0,0,even),  (odd,0,0,0,odd), (0,odd,0,0,even), (0,odd,0,0,odd) in modes (7,8,9,10,5), respectively. For these cases, the teleportation is unidirectional in a sense that Bob can get the information of Alice (with fidelity either unity or $F_{3}^{AB}$) but Alice's state is either an even or odd coherent state depending upon the outcome of the controller Charlie. 

For cases 3 to 6, the received states with \{Alice,Bob\} are given by, $\{|A^{0}\rangle,|I^{B}\rangle\}$, $\{|A^{1}\rangle,|B^{3}\rangle\}$, $\{|A^{0}\rangle,|B^{2}\rangle\}$, $\{|A^{1}\rangle,-|B^{4}\rangle\}$, respectively, requiring unitary operations,  $\{I,I\}$, $\{U_{1},U_{1}\}$, $\{I,U_{3}\}$ and $\{U_{1},-U_{2}\}$, respectively (where $I$ is the identity operation). Although, performing non-identity operation by Alice will not help her to get to any close to Bob's information, however, it does later help us to increase the average fidelity of our scheme. The fidelity of teleported state with Alice, $F^{BA}$, becomes $F^{BA}_{1}$ for cases 3 to 6, while the fidelity of Bob, $F^{AB}$, is unity for cases 3 and 5, and $F^{AB}_{3}$ for cases 4 and 6. It can be observed by Eq. \ref{eqn:8} that the teleported state, the required unitary operation and corresponding fidelities for cases 7, 8, 9 and 10 are same as that for cases 4, 3, 6 and 5 respectively.

The probability of obtaining case 3 (same for case 5) is given by
\begin{equation}
\label{eqn:15}
P_{II,+}=\frac{x^2(1-x^2)^{2}(1+x^{2})|B_{+}|^{2}}{4(1+2x^{6}+x^{8})}
\end{equation}
while the probability of obtaining case 4 (same for case 6) is given by
\begin{equation}
\label{eqn:16} 
P_{II,-}=\frac{x^2(1-x^2)^{4}(\gamma^{2}|A_{+}|^{2}+\delta^{2}|A_{-}|^{2})|B_{+}|^{2}}{4(1+2x^{6}+x^{8})}.
\end{equation}
Similarly, the probability of obtaining case 7 (same for case 9) is given by,
\begin{equation}
\label{eqn:17} 
P_{III,+}=\frac{x^2(1+x^2)^{2}(1-x^2)(\gamma^{2}|A_{+}|^{2}+\delta^{2}|A_{-}|^{2})|B_{+}|^{2}}{4(1+2x^{6}+x^{8})},
\end{equation}
while for case 8 (same for case 10) is given by
\begin{equation}
\label{eqn:18} 
P_{III,-}=\frac{x^2(1-x^2)^{3}|B_{+}|^{2}}{4(1+2x^{6}+x^{8})}.
\end{equation}
The probabilities $P_{III,-}$ and $P_{IV,+}$ are function of $x$ and information parameters $\theta$ and $\theta^{'}$ while $P_{III,+}$ and $P_{IV,-}$ only depend on $x$ and $\theta^{'}$. We have plotted these probabilities in units of $|B_{+}|^2$ in Fig. \ref{fig:3} from which it is evident that the probabilities decrease with increasing coherent amplitude to become almost zero. Hence, one can infer that these cases also have little chance of occurrence at appreciably large coherent amplitude ($|\alpha|^{2}\geq 3$).

\begin{figure}[t]
  \includegraphics[width=\linewidth]{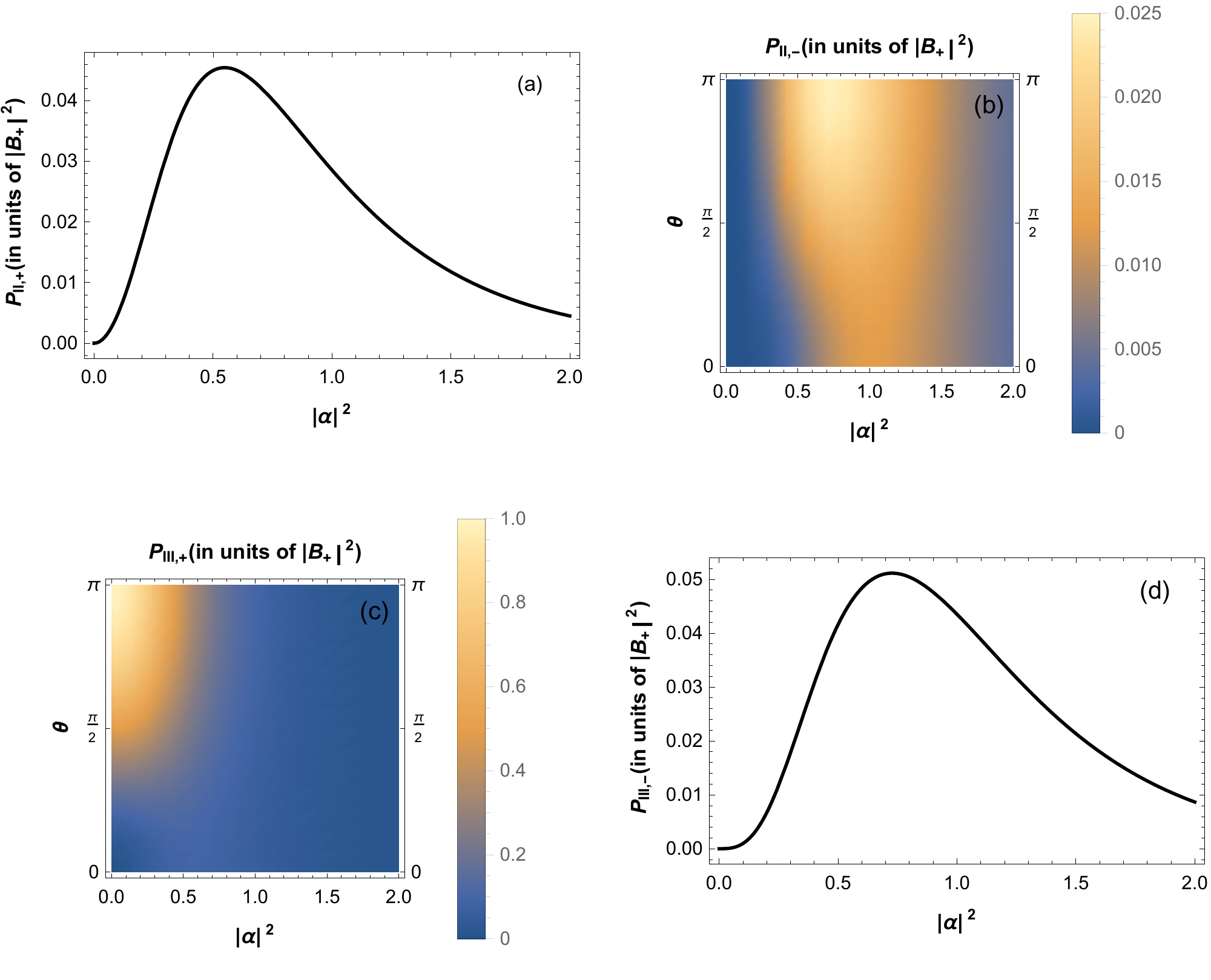}
  \caption{(a), (b), (c) and (d) show the variation of probability of occurrence in units of $|B_{+}|^{2}$ for cases 3, 4, 7 and 8 with respect to $|\alpha|^{2}$ and information parameter $\theta$. The probabilities becomes almost zero for $|\alpha|^{2} \geq 2$.}
  \label{fig:3}
\end{figure}
  
\subsubsection{11 to 18}
Cases 11 to 18 occur for collective PC outcomes (0,0,NZE,0,even), (0,0,NZE,0,odd), (0,0,0,NZE,even), (0,0,0,NZE,odd), (0,0,odd,0,even), (0,0,odd,0,odd), (0,0,0,odd,even), (0,0,0,odd,odd) in modes (7,8,9,10,5), respectively. These cases can be obtained by inter-changing the roles of Alice and Bob in cases 3 to 10, respectively. Therefore, the received state with Alice (Bob) for cases 11 to 18 will be that of Bob (Alice) for cases 3 to 10, with $B_{\pm} (A_{\pm})$ replaced with $A_{\pm} (B_{\pm})$, respectively. Also, the expression for probability for these cases can be obtained using eq. 16 to 19, by the same replacement of $A(B)$ to $B(A)$. Explicitly, if $P_{IV,+}$,  $P_{IV,-}$, $P_{V,+}$ and  $P_{V,-}$ represent the probability of obataining case 11 (same for case 13), case 12 (same for case 14), case 15 (same for 17) and case 16 (same for case 18), respectively, we have,
\begin{equation}
\label{eqn:19} 
P_{IV,+}=\frac{x^2(1-x^2)^{2}(1+x^{2})|A_{+}|^{2}}{4(1+2x^{6}+x^{8})},
\end{equation}
\begin{equation}
\label{eqn:20} 
P_{IV,-}=\frac{x^2(1-x^2)^{4}(\gamma^{2}|B_{+}|^{2}+\delta^{2}|B_{-}|^{2})|A_{+}|^{2}}{4(1+2x^{6}+x^{8})},
\end{equation}
\begin{equation}
\label{eqn:21} 
P_{V,+}=\frac{x^2(1+x^2)^{2}(1-x^2)(\gamma^{2}|B_{+}|^{2}+\delta^{2}|B_{-}|^{2})|A_{+}|^{2}}{4(1+2x^{6}+x^{8})},
\end{equation}
\begin{equation}
\label{eqn:22} 
P_{V,-}=\frac{x^2(1-x^2)^{3}|A_{+}|^{2}}{4(1+2x^{6}+x^{8})}.
\end{equation}
If we plot these probabilities in units of $|A_{+}|^{2}$, we would have similar variation as we have already obtained for expression 16 to 19 as shown in Fig. \ref{fig:2}. 

\subsection{Cases of bi-directional teleportation}
\subsubsection{Cases 19 to 34}
Cases 19 to 34 correspond to collective PC outcome (NZE,0,NZE,0,even), (NZE,0,NZE,0,odd), (NZE,0,0, NZE,even), (NZE,0,0,NZE,odd), (0,NZE,NZE,0,even), (0,NZE,NZE,0,odd), (0,NZE,0,NZE,even), (0,NZE, 0,NZE,odd), (odd,0,odd,0,even), (odd,0,odd,0,odd), (odd,0,0,odd,even), (odd,0,0,odd,odd), (0,odd,odd,0 ,even), (0,odd,odd,0,odd), (0,odd,0,odd,even), (0,odd,0,odd,odd) in modes (7,8,9,10,5), respectively. We have for these cases, the received state with \{Alice,Bob\} as, $\{|I^{B}\rangle,|I^{A}\rangle\}$, $\{|A^{3}\rangle,|B^{3}\rangle\}$, $\{|A^{2}\rangle,|I^{A}\rangle\}$, $\{|A^{4}\rangle,|B^{3}\rangle\}$, $\{|I^{B}\rangle,|B^{2}\rangle\}$, $\{|A^{4}\rangle,|B^{4}\rangle\}$, $\{|A^{2}\rangle,|B^{2}\rangle\}$, $\{|A^{4}\rangle,|B^{4}\rangle\}$, $\{|A^{3}\rangle,|B^{3}\rangle\}$, $\{|I^{B}\rangle,|I^{A}\rangle\}$, $\{|A^{3}\rangle,|B^{3}\rangle\}$, $\{|A^{2}\rangle,|I^{A}\rangle\}$, $\{|A^{3}\rangle,|B^{4}\rangle\}$, $\{|I^{B}\rangle,|B^{2}\rangle\}$, $\{|A^{4}\rangle,|B^{4}\rangle\}$, $\{|A^{2}\rangle,|B^{2}\rangle\}$, respectively, requiring unitary operations, $\{I,I\}$, $\{U_{1},U_{1}\}$, $\{U_{3},I\}$ and $\{U_{2},U_{1}\}$, $\{I,U_{3}\}$, $\{U_{2},U_{2}\}$, $\{U_{3},U_{3}\}$, $\{U_{2},U_{2}\}$, $\{U_{1},U_{1}\}$, $\{I,I\}$, $\{U_{1},U_{1}\}$, $\{U_{3},I\}$, $\{U_{1},U_{2}\}$, $\{I,U_{2}\}$, $\{U_{2},U_{2}\}$, $\{U_{3},U_{3}\}$, respectively. For all these cases, there is either perfect teleportation (in both way) with unit fidelity or with fidelity almost unity at appreciable coherent amplitude. The fidelity of state teleported to Alice (Bob) then becomes unity for cases 19, 21, 23, 25, 28, 30, 32, 34 and $F_{3}^{BA}$($F_{3}^{AB}$) for cases 20, 22, 24, 26, 27, 29, 31 and 33.

Representing $P_{VI,+}$ and $P_{VI,-}$ the probability of obtaining case 19 (same for case 21, 23, 25) and case 20 (same for 22,24 and 26), respectively, are given by
\begin{equation}
\label{eqn:23} 
P_{VI,+}=\frac{(1-x^{2})^{4}(1+x^{2})}{32(1+2x^{6}+x^{8})}
\end{equation} 
\begin{equation}
\label{eqn:24} 
P_{VI,-}=\frac{(1-x^{2})^{5}(\gamma^{2}|A_{+}|^{2}+\delta^{2}|A_{-}|^{2})(\gamma^{2}|B_{+}|^{2}+\delta^{2}|B_{-}|^{2})}{32(1+2x^{6}+x^{8})}.
\end{equation}
Similarly, if $P_{VII,+}$ and $P_{VII,-}$ represent the probability of obtaining case 27 (same for case 29, 31, 33) and case 28 (same for 30, 32 and 34), respectively, then we have
\begin{equation}
\label{eqn:25} 
P_{VII,+}=\frac{(1-x^{2})^{2}(1+x^{2})^{3}(\gamma^{2}|A_{+}|^{2}+\delta^{2}|A_{-}|^{2})(\gamma^{2}|B_{+}|^{2}+\delta^{2}|B_{-}|^{2})}{32(1+2x^{6}+x^{8})}
\end{equation} 
\begin{equation}
\label{eqn:26} 
P_{VII,-}=\frac{(1-x^{2})^{3}(1+x^{2})^{2}}{32(1+2x^{6}+x^{8})}.
\end{equation}
For $|\alpha|^{2}\geq 2$, $x\simeq 0$, implying, $\gamma,\delta\rightarrow 1$. In this particular limit, we have $P_{VI,\pm}$ and $P_{VII,\pm}$ asymptotically converges to a constant value of 1/32.

\subsubsection{35 to 50}
Cases 35 to 42 correspond to collective PC outcome (NZE,0,odd,0,even), (NZE,0,odd,0,odd), (NZE,0,0, odd,even), (NZE,0,0,odd,odd), (0,NZE,odd,0,even), (0,NZE,odd,0,odd), (0,NZE,0, odd,even), (0,NZE,0, odd,odd), (odd,0,NZE,0,even), (odd,0,NZE,0,odd), (odd,0,0,NZE,even), (odd,0,0,NZE,odd), (0,odd,NZE, 0,even), (0,odd,NZE,0,odd), (0,odd,0,NZE,even), (0,odd,0,NZE,odd) in modes (7,8,9,10,5), respectively. For these cases, the received state with \{Alice,Bob\} are given by $\{|A^{3}\rangle,|I^{A}\rangle\}$, $\{|I^{B}\rangle,|B^{3}\rangle\}$, $\{|A^{4}\rangle,|I^{A}\rangle\}$, $\{|I^{B}\rangle,|B^{3}\rangle\}$, $\{|A^{3}\rangle,|B^{2}\rangle\}$, $\{|I^{B}\rangle,|B^{4}\rangle\}$, $\{|A^{3}\rangle,|B^{2}\rangle\}$, $\{|A^{2}\rangle,-|B^{4}\rangle\}$
, $\{|I^{B}\rangle,|B^{3}\rangle\}$, $\{|A^{3}\rangle,|I^{A}\rangle\}$, $\{|A^{2}\rangle,|B^{3}\rangle\}$, $\{|A^{4}\rangle,|I^{A}\rangle\}$, $\{|I^{B}\rangle,-|B^{4}\rangle\}$, $\{|A^{3}\rangle,-|B^{2}\rangle\}$, $\{|A^{2}\rangle,-|B^{4}\rangle\}$, $\{-|A^{4}\rangle,|B^{2}\rangle\}$, respectively, requiring unitary operations,  $\{U_{1},I\}$, $\{I,U_{1}\}$, $\{U_{2},I\}$, $\{I,U_{1}\}$, $\{U_{1},U_{2}\}$, $\{I,U_{2}\}$, $\{U_{1},U_{3}\}$, $\{U_{3},-U_{2}\}$, $\{I,U_{1}\}$, $\{U_{1},I\}$, $\{U_{3},U_{1}\}$, $\{U_{2},I\}$, $\{I,-U_{2}\}$, $\{U_{1},-U_{3}\}$, $\{U_{3},-U_{2}\}$, $\{-U_{2},U_{3}\}$, respectively. The fidelity of state teleported to Alice and Bob becomes 1 and $F_{3}^{AB}$, respectively, for cases 36, 38, 40, 42, 43, 45, 47 and 49. For rest of the cases, 37, 39, 41, 44, 46, 48 and 50, the fidelity becomes $F_{3}^{BA}$ for teleported state with Alice, and unity for state teleported to Bob.

The probability of obtaining case 35 (same for case 37, 39, 41) becomes
\begin{equation}
\label{eqn:27} 
P_{VIII,+}=\frac{(1-x^{2})^{3}(1+x^{2})^{2}(\gamma^{2}|B_{+}|^{2}+\delta^{2}|B_{-}|^{2})}{32(1+2x^{6}+x^{8})},
\end{equation}
while probability of obtaining case 36 (same for case 38, 40, 42) is
\begin{equation}
\label{eqn:28} 
P_{VIII,-}=\frac{(1-x^{2})^{4}(1+x^{2})(\gamma^{2}|A_{+}|^{2}+\delta^{2}|A_{-}|^{2})}{32(1+2x^{6}+x^{8})}
\end{equation}
Similarly, the probability of obtaining case 43 (same for case 45, 47, 49) is given by
\begin{equation}
\label{eqn:29} 
P_{IX,+}=\frac{(1-x^{2})^{3}(1+x^{2})^{2}(\gamma^{2}|A_{+}|^{2}+\delta^{2}|A_{-}|^{2})}{32(1+2x^{6}+x^{8})}
\end{equation}
and the probability of obtaining case 44 (same for case 46, 48, 50) becomes as
\begin{equation}
\label{eqn:30} 
P_{IX,-}=\frac{(1-x^{2})^{4}(1+x^{2})(\gamma^{2}|B_{+}|^{2}+\delta^{2}|B_{-}|^{2})}{32(1+2x^{6}+x^{8})}.
\end{equation}
The probabilities $P_{VIII,\pm}$ and $P_{IX,\pm}$ also converge asymptotically to 1/32 for $|\alpha|^{2}\geq 2$.
 
\section{Average fidelity and minimum assured fidelity of teleportation }
\begin{figure}[t]
  \includegraphics[width=\linewidth]{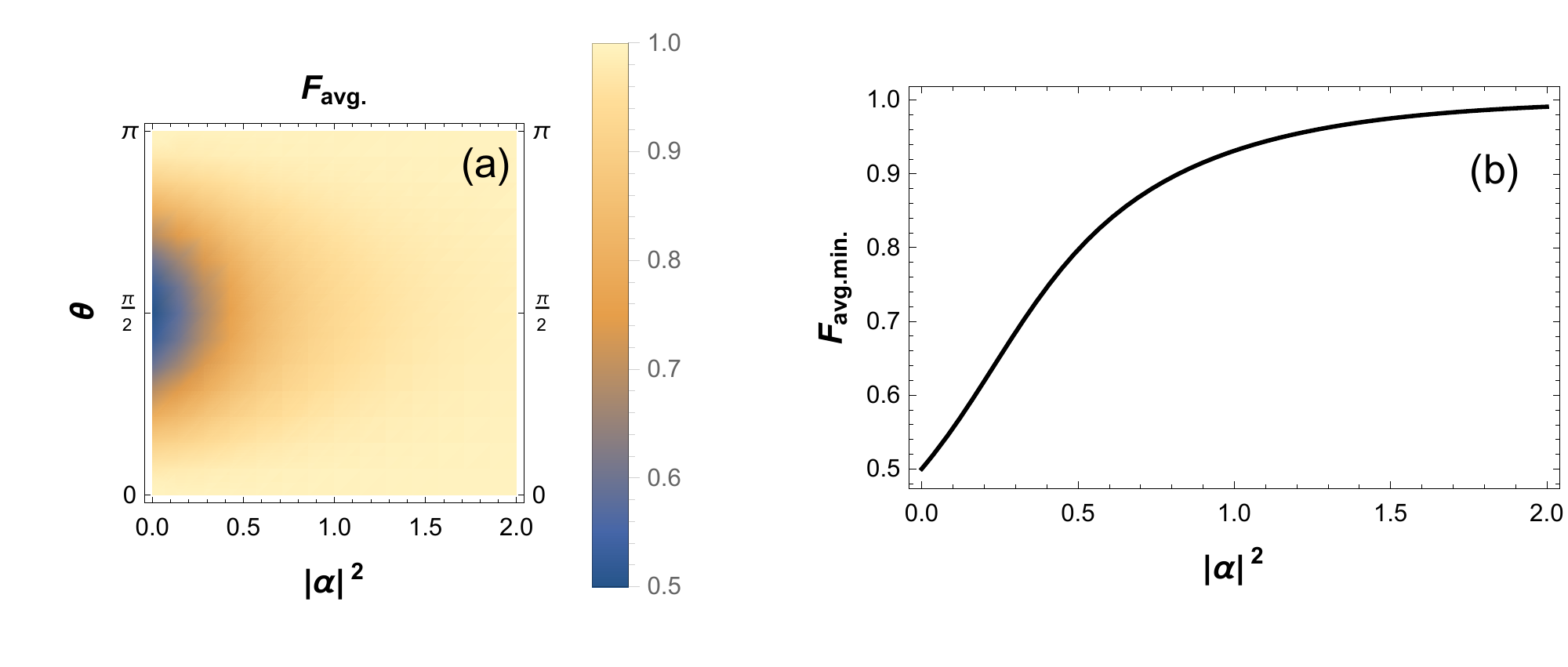}
  \caption{(a) Variation of average fidelity of teleportation with respect to information parameter $\theta$ and mean photons in the coherent state, $|\alpha|^{2}$. (b) Variation of minimum average fidelity of teleportation with respect to mean photons in the coherent state, $|\alpha|^{2}$. Both increase monotonically to become almost unity for $|\alpha|^{2}\geq 2$.}
  \label{fig:3}
\end{figure}
In view of previous section, it is evident that the individual fidelities $F^{AB}$ and $F^{BA}$ for many cases depend on information parameter, $\theta$ and $\theta^{'}$ as well as on mean coherent amplitude $|\alpha|^{2}$. In such a scenario to effectively quantify  the overall quality of teleportation, we find the average fidelity of teleportation ($F_{avg.}$) which is the sum of product of individual fidelity with its probability of occurrence. For individual cases, we calculate the minimum assured fidelity (MAF) which is the minimum value of fidelity for every possible information state \cite{prakash2012minimum}.

As there are two teleportations involved, we need to find average fidelity for both teleportation from Alice to Bob, denoted by $F_{avg.}^{AB}$ and $F^{BA}_{avg.}$ for teleportation from Bob to Alice. Although, $F^{AB}\neq F^{BA}$ for several cases, it turns out that $F_{avg.}^{AB}(F^{BA}_{avg.})$ have similar functional dependence on information parameters as well as $x$,
\begin{equation}
\label{eqn:31} 
F^{AB(BA)}_{avg.}\sum_{i=1}^{50}P_{i}F_{i}=1-\frac{x^{2}(1-x^{2}+2x^{4})\sin^{2}\theta(\theta^{'})}{2(1-x^{2}+x^{4}+x^{6})}.
\end{equation}
$F^{AB(BA)}_{avg.}$ minimizes for $\theta(\theta^{'})=\pi/2$, while its maxima is obtained for $\theta(\theta^{'})=\{0,\pi\}$ equal to 1. The average fidelity $F^{AB}_{avg.}$ as well as $F^{AB}_{avg.min.}$ is plotted with respect to information parameter $\theta$ and $|\alpha|^{2}$ in Fig.\ref{fig:3}. It is evident that both becomes almost unity for $|\alpha|^{2}\geq 2$.

We shall now discuss MAF for all PC cases.  Let us first consider those cases for which the fidelity of state teleported to either Bob or Alice is one of $F_{1}^{AB}, F_{2}^{AB}$, $F_{1}^{BA}, F_{2}^{BA}$.  For all such cases, the fidelity is proportional to either $|A_{\pm}|$ or $|B_{\pm}|$, which is zero if information states are $\theta=\pi/2,0$ or $\theta^{'}=\pi/2,0$, respectively, and for all allowed value of $\phi$ and $\phi^{'}$. Hence for these cases, MAF is 0. For those cases in which the teleported state with Alice or Bob gives fidelity, $F_{3}^{AB}$ or $F_{3}^{BA}$, then a minimum is obtained for $\theta^{(')}=\cos^{-1}x^{2}$, therefore, MAF for such cases becomes, $(1-x^{4})$. For rest of the cases, fidelity is unity and we have ideal teleportation. 
\section{Conclusion}
We may conclude that CBQT of information encoded in phase opposite state coherent state can be obtained using five-mode cluster-type ECS as a resource with average fidelity of almost unity at moderate coherent amplitude. We note that complete failure of CBQT is obtained for cases 1 and 2, while uni-directional teleportation is obtained for cases 3 to 18. However, the probability of occurrence of all such cases becomes negligibly small as we increase mean photon in coherent state. For each of Cases 19 to 50, we obtain almost perfect success for CBQT in terms of fidelity of teleportation. The probability of occurrence of each of these cases becomes almost equal converging asymptotically to a constant value of 1/32 for $|\alpha|^{2}\geq 2$. We only require linear optical devices, such as, symmetric beam splitters, phase shifters, and photon counters for its practical realization. 
\section*{Acknowledgement}
We dedicate this paper in the memory of Prof. Ranjana Prakash and Prof. Hari Prakash. May their soul rest in peace. RKP is thankful to UGC for providing financial assistance under CSIR-UGC SRF fellowship. Discussions with Dr. D K Mishra, Dr. M K Mishra, Dr. V Verma and Ms. S Javed are gratefully acknowledged.
\appendix
\section{Appendix}
\begin{longtable}[h!]{ c c c c c c c c c c c }
\hline 
\label{my-label}
 \parbox[t]{2mm}{\multirow{2}{*}{\rotatebox[origin=c]{90}{Cases}}} & \multicolumn{3}{c}{PC result obtained by }  & \parbox[t]{2mm}{\multirow{2}{*}{\rotatebox[origin=c]{90}{Probability}}} &   \multicolumn{2}{c}{\thead{Unitary\\ operation}}& \multicolumn{2}{c}{\thead{Teleported  \\ state}} & \multicolumn{2}{c}{Fidelity}   
\\
 & \thead{Alice \\ (7,8)}  &
  \thead{Bob \\ (9,10)} 
  & \thead{Charlie\\ (5)} &  & 
  Alice & Bob & Alice & Bob & Alice & Bob  \\
   \hline 
   & & & & & & & & & &\\
  1 & (0,0)  & (0,0) & even & $P_{I,+}$ & $I$ & $I$  & $|A^{0}\rangle $ & $|B^{0}\rangle $ & $F_{1}^{AB}$ & $F_{1}^{BA}$ 
  \\
  2 & (0,0)  & (0,0) & odd & $P_{I,-}$ &  $I$ & $I$  & $|A^{1}\rangle $ & $|B^{1}\rangle $ & $F_{2}^{AB}$ & $F_{2}^{BA}$  \\
\hline
  3 & (NZE,0)  & (0,0) & even & $P_{II,+}$ &  $I$ & $I$ & $|A^{0}\rangle$ & $|I^{B}\rangle $  & $F_{1}^{AB}$ & 1   
   \\
 4 & (NZE,0)  & (0,0) & odd & $P_{II,-}$ & $U_{1}$ & $U_{1}$  & $|A^{0}\rangle $ & $|I^{'A}\rangle $ &$F_{1}^{AB}$ & $F_{3}^{BA}$    \\ \hline
 5 & (0,NZE)  & (0,0) & even & $P_{II,+}$ & $I$ & $U_{3}$  & $|A^{0}\rangle $ & $|I^{A}\rangle $ & $F_{1}^{AB}$ & 1 
 \\
 6 & (0,NZE)  & (0,0) & odd & $P_{II,-}$ &  $U_{1}$ & $U_{2}$ & $|A^{0}\rangle $ & $|I^{'A}\rangle $ & $F_{1}^{AB}$ & $F_{3}^{BA}$   
 \\
\hline
7 & (odd,0)  & (0,0) & even & $P_{III,+}$ & $I$ & $U_{1}$   & $|A^{0}\rangle $ & $|I^{'A}\rangle $ & $F_{1}^{AB}$ & $F_{3}^{BA}$   \\ 
8 & (odd,0)  & (0,0) & odd & $P_{III,-}$ & $U_{1}$ & $I$  & $|A^{0}\rangle $ & $|I^{A}\rangle $ & $F_{1}^{AB}$ & 1 \\
\hline
9 & (0,odd)  & (0,0) & even & $P_{III,+}$ & $I$ & $U_{2}$ & $|A^{0}\rangle $ &$|I^{'A}\rangle $ & $F_{1}^{AB}$ & $F_{3}^{BA}$     \\ 
10 & (0,odd)  & (0,0) & odd & $P_{III,-}$ & $U_{1}$ & $U{3}$ & $|A^{0}\rangle $ & $|I^{A}\rangle $  & $F_{1}^{AB}$ & $F_{3}^{BA}$        \\
\hline
11 & (0,0)  & (NZE,0) & even & $P_{IV,+}$ & $\hat{I}$ & $\hat{I}$ & $|I^{B}\rangle $ & $|B^{0}\rangle $  & 1 & $F_{1}^{BA}$   
 \\
12 & (0,0)  & (NZE,0) & odd & $P_{IV,-}$ & $U_{1}$ & $U_{1}$ & $|I^{'B}\rangle $ & $|B^{0}\rangle $  & $F_{3}^{AB}$ & $F_{1}^{BA}$    \\
\hline

13 & (0,0)  & (0,NZE) & even & $P_{IV,+}$ & $U_{3}$ & $I$ & $|I^{B}\rangle $ & $|B^{0}\rangle $  & 1 & $F_{1}^{BA}$   
 \\
14 & (0,0)  & (0,NZE) & odd & $P_{IV,-}$ & $U_{2}$ & $U_{1}$ & $|I^{'B}\rangle $ & $|B^{0}\rangle $  & $F_{3}^{AB}$ & $F_{1}^{BA}$    \\
\hline

15 & (0,0)  & (odd,0) & even & $P_{V,+}$ & $U_{1}$ & $\hat{I}$   &$|I^{'B}\rangle $ & $B^{0}\rangle $ & $F_{3}^{AB}$ & $F_{1}^{BA}$       \\

16 & (0,0)  & (odd,0) & odd & $P_{V,-}$ & $I$ & $U_{1}$ &  $|I^{B}\rangle $ & $|B^{0}\rangle $ & 1 & $F_{1}^{BA}$    \\
\hline

17 & (0,0)  & (0,odd) & even & $P_{V,+}$ & $U_{2}$ & $\hat{I}$   &$|I^{'B}\rangle $ & $B^{0}\rangle $ & $F_{3}^{AB}$ & $F_{1}^{BA}$       \\

18 & (0,0)  & (0,odd) & odd & $P_{V,-}$ & $U_{3}$ & $U_{1}$ &  $|I^{B}\rangle $ & $|B^{0}\rangle $ & 1 & $F_{1}^{BA}$    \\
\hline

19 & (NZE,0)  & (NZE,0) & even & $P_{VI,+}$ & $I$ & $I$  & $|I^{B}\rangle $ & $|I^{A}\rangle $ & 1 & 1    \\

20 & (NZE,0)  & (NZE,0) & odd & $P_{VI,-}$  & $U_{1}$ & $U_{1}$  & $|I^{'B}\rangle $ & $|I^{'A}\rangle $ & $F_{3}^{AB}$ & $F_{3}^{BA}$  \\
\hline

21 & (NZE,0)  & (0,NZE) & even & $P_{VI,+}$ & $U_{3}$ & $U_{3}$  & $|I^{B}\rangle $ & $|I^{A}\rangle $ & 1 & 1  \\

22 & (NZE,0)  & (0,NZE) & odd & $P_{VI,-}$ & $U_{2}$ & $U_{2}$  & $|I^{'B}\rangle $ & $|I^{'A}\rangle $ & $F_{3}^{AB}$ & $F_{3}^{BA}$   \\
\hline

23 & (0,NZE)  & (NZE,0) & even & $P_{VI,+}$ & $I$ & $U_{3}$  & $|I^{B}\rangle $ & $|I^{A}\rangle $ & 1 & 1     \\
\hline
24 & (0,NZE)  & (NZE,0) & odd & $P_{VI,-}$ & $U_{1}$ & $U_{2}$  & $|I^{'B}\rangle $ & $|I^{'A}\rangle $ & $F_{3}^{AB}$ & $F_{3}^{BA}$ \\
\hline

25 & (0,NZE)  & (0,NZE) & even & $P_{VI,+}$& $U_{3}$ & $U_{3}$  & $|I^{B}\rangle $ & $|I^{A}\rangle $ & 1 & 1 \\

26 & (0,NZE)  & (0,NZE) & odd & $P_{VI,-}$ & $U_{2}$ & $U_{2}$  & $|I^{'B}\rangle $ & $|I^{'A}\rangle $ & $F_{3}^{AB}$ & $F_{3}^{BA}$           \\
\hline

27 & (odd,0)  & (odd,0) & even & $P_{VII,+}$ & $U_{1}$ & $U_{1}$  & $|I^{'B}\rangle $ & $|I^{'A}\rangle $ & $F_{3}^{AB}$ & $F_{3}^{BA}$  \\

28 & (odd,0)  & (odd,0) & odd & $P_{VII,-}$ & $I$ & $I$  & $|I^{B}\rangle $ & $|I^{A}\rangle $ & 1 & 1    \\
\hline

29 & (odd,0)  & (0,odd) & even & $P_{VII,+}$ & $U_{1}$ & $U_{1}$  & $|I^{'B}\rangle $ & $|I^{'A}\rangle $ & $F_{3}^{AB}$ & $F_{3}^{BA}$  \\

30 & (odd,0)  & (0,odd) & odd & $P_{VII,-}$ & $U_{3}$ & $I$  & $|I^{B}\rangle $ & $|I^{A}\rangle $ & 1 & 1   \\
\hline

31 & (0,odd)  & (odd,0) & even & $P_{VII,+}$ & $U_{1}$ & $U_{2}$  & $|I^{'B}\rangle $ & $|I^{'A}\rangle $ & $F_{3}^{AB}$ & $F_{3}^{BA}$    \\

32 & (0,odd)  & (odd,0) & odd & $P_{VII,-}$ & $I$ & $U_{3}$  & $|I^{B}\rangle $ & $|I^{A}\rangle $ & 1 & 1  \\
\hline

33 & (0,odd)  & (0,odd) & even & $P_{VII,+}$ & $U_{2}$ & $U_{2}$  & $|I^{'B}\rangle $ & $|I^{'A}\rangle $ & $F_{3}^{AB}$ & $F_{3}^{BA}$    \\

34 & (0,odd)  & (0,odd) & odd & $P_{VII,-}$ & $U_{3}$ & $U_{3}$  & $|I^{B}\rangle $ & $|I^{A}\rangle $ & 1 & 1 \\
\hline

35 & (NZE,0)  & (odd,0) & even & $P_{VIII,+}$ & $U_{1}$ & $I$  & $|I^{'B}\rangle $ & $|I^{A}\rangle $ & $F_{3}^{AB}$ & 1    \\

36 & (NZE,0)  & (odd,0) & odd & $P_{VIII,-}$ & $I$ & $U_{1}$  & $|I^{B}\rangle $ & $|I^{'A}\rangle $ & 1 & $F_{3}^{BA}$   \\
\hline

37 & (NZE,0)  & (0,odd) & even & $P_{VIII,+}$ & $U_{2}$ & $I$  & $|I^{'B}\rangle $ & $|I^{A}\rangle $ & $F_{3}^{AB}$ & 1   \\

38 & (NZE,0)  & (0,odd) & odd & $P_{VIII,-}$ & $I$ & $U_{1}$  & $|I^{B}\rangle $ & $|I^{'A}\rangle $ & 1 & $F_{3}^{BA}$    \\
\hline

39 & (0,NZE)  & (odd,0) & even & $P_{VIII,+}$ & $U_{1}$ & $U_{3}$  & $|I^{'B}\rangle $ & $|I^{A}\rangle $ & $F_{3}^{AB}$ & 1 \\ 

40 & (0,NZE)  & (odd,0) & odd & $P_{VIII,-}$  & $I$ & $U_{2}$  & $|I^{B}\rangle $ & $|I^{'A}\rangle $ & 1 & $F_{3}^{BA}$ \\
\hline

41 & (0,NZE)  & (0,odd) & even & $P_{VIII,+}$ & $U_{1}$ & $U_{3}$  & $|I^{'B}\rangle $ & $|I^{A}\rangle $ & $F_{3}^{AB}$ & 1 \\

42 & (0,NZE)  & (0,odd) & odd & $P_{VIII,-}$  & $U_{3}$ & $-U_{2}$  & $|I^{B}\rangle $ & $|I^{'A}\rangle $ & 1 & $F_{3}^{BA}$     \\
\hline

43 & (odd,0)  & (NZE,0) & even & $P_{IX,+}$  & $I$ & $U_{1}$  & $|I^{B}\rangle $ & $|I^{'A}\rangle $ & 1 & $F_{3}^{BA}$ \\

44 & (odd,0)  & (NZE,0) & odd & $P_{IX,+}$ & $U_{1}$ & $I$  & $|I^{'B}\rangle $ & $|I^{A}\rangle $ & $F_{3}^{AB}$ & 1   \\
\hline

45 & (odd,0)  & (0,NZE) & even & $P_{IX,+}$ & $U_{3}$ & $U_{1}$  & $|I^{B}\rangle $ & $|I^{'A}\rangle $ & 1 & $F_{3}^{BA}$   \\

46 & (odd,0)  & (0,NZE) & odd & $P_{IX,+}$ & $U_{2}$ & $I$  & $|I^{'B}\rangle $ & $|I^{A}\rangle $ & $F_{3}^{AB}$ & 1  \\
\hline

47 & (0,odd)  & (NZE,0) & even & $P_{IX,+}$ & $I$ & $-U_{2}$  & $|I^{B}\rangle $ & $|I^{'A}\rangle $ & 1 & $F_{3}^{BA}$ \\

48 & (0,odd)  & (NZE,0) & odd & $P_{IX,+}$ & $U_{1}$ & $-U_{3}$  & $|I^{'B}\rangle $ & $|I^{A}\rangle $ & $F_{3}^{AB}$ & 1     \\
\hline
49 & (0,odd)  & (0,NZE) & even & $P_{IX,+}$ & $U_{3}$ & $-U_{2}$  & $|I^{B}\rangle $ & $|I^{'A}\rangle $ & 1 & $F_{3}^{BA}$     \\

50 & (0,odd)  & (0,NZE) & odd & $P_{IX,+}$ & -$U_{2}$ & $U_{3}$  & $|I^{'B}\rangle $ & $|I^{A}\rangle $ & $F_{3}^{AB}$ & 1      \\
\hline
\caption{Various possible PC measurement results in modes 7, 8 with Alice, modes 9, 10 with Bob, and mode 5 with controller Charlie. For each case, the corresponding probability of occurrence, required unitary operation, teleported state and fidelity is given. }
\label{table1}
\end{longtable}
\renewcommand{\arraystretch}{2}

\bibliographystyle{unsrt}

\end{document}